%% file: smtt.tex
\pdfoutput=1
\pdfmapfile{+mlm.map}

\documentclass{scrartcl}

\usepackage{preamble}
\usepackage{tikz}
\usepackage{pftools}
\usepackage{bbold}
\usepackage{mtt-macros}
\usepackage{authblk}

\title{Under Lock and Key: \\ A Proof System for a Multimodal Logic%
\footnote{
  ACM CCS \quad Theory of computation: Modal and temporal logics, Proof theory, Type theory
}}

\author[1]{G. A. Kavvos}
\author[2]{Daniel Gratzer}
\affil[1]{School of Computer Science, University of Bristol}
\affil[2]{Department of Computer Science, Aarhus University}

\date{\small 17 April 2023}

\begin{document}

\maketitle

\begin{abstract}
  We present a proof system for a multimode and multimodal logic, which is based on our
  previous work on modal Martin-L\"{o}f type theory. The specification of
  modes, modalities, and implications between them is given as a mode theory,
  i.e. a small 2-category. The logic is extended to a lambda calculus,
  establishing a Curry-Howard correspondence.
\end{abstract}

\section{Introduction}

Many-dimensional~\parencite{gabbay_2003}, multimodal~\parencite{carnielli_2008}
or poly-modal~\parencite{benthem_2010} logics have found a number of successful
applications. To name but a few:
\begin{center}
  \begin{tabular}{ccc}
    temporal logic & \(\mathbf{F}\varphi\), \(\mathbf{G}\varphi\), \(\mathbf{X}\varphi\) & \parencite{demri_2016} \\
    epistemic logic & \(K_i\varphi\), \(B_i\varphi\), \(C_G\varphi\)  & \parencite{fagin_1995} \\
    dynamic logic & \([a]\varphi\), \(\langle a \rangle\varphi\) & \parencite{harel_2000} \\
    dynamic epistemic logic & \(K_i\varphi\), \([\alpha]\varphi\) & \parencite{van_ditmarsch_2008}  \\
    Hennessy-Milner logic & \([\alpha]\varphi\), \(\langle \alpha \rangle\varphi\) & \parencite{stirling_2001}
  \end{tabular}
\end{center}
The majority of work on the aforementioned logics has a number of common features:
\begin{itemize}
  \item \textbf{The propositional substrate is almost always classical.} While a
    classical approach is more than sufficient for modelling knowledge and
    computational systems, it precludes the making of a close connection with
    categorical logic, where the \emph{internal language} of many categories is intuitionistic
    \parencite{pitts_2001}.

  \item \textbf{The modal fragment is almost always inspired by a Kripke
    semantics, and lacks a proof system.} The Kripke semantics usually model
    some intensional aspect of interest, such as states of knowledge, the
    execution trace of a machine, and so on. While this is indeed more than
    adequate for modelling purposes, it precludes the immediate formulation of a
    well-behaved, computational theory for these logics under the
    Curry-Howard correspondence \parencite{girard_1989, sorensen_2006}.

  \item \textbf{There is no cohesive, unifying account.} While there have been a
    few attempts at building a framework \parencite[\S 8]{carnielli_2008}, as
    well as a host of results on combining simpler modal logics using
    \emph{product} and \emph{fusion} operators
    \parencite[\S\S3--4]{gabbay_2003}, we have yet to obtain a unifying account
    of logics with multiple interacting modalities.
\end{itemize}

In this paper we present a new modal logic. Unlike previous work, this logic
fixes neither the number nor the interactions of modalities in advance. Instead,
it is given parametrically in a specification of the modalities and their
interrelations, which is called the \emph{mode theory}.

Moreover, this new logic is not just \emph{multimodal}---in that it sports
multiple modalities---but also \emph{multimode}. This is a new concept in modal
logic. Traditionally, a modal operator \(\Box\) is an operator that takes a
formula \(\varphi\) to a formula \(\Box \varphi\). Crucially, the formula
\(\Box\varphi\) is in the same syntactic category as \(\varphi\). The logic in
this paper will conceive of modal operators as transporting formulas
\emph{between} multiple syntactic categories. We will call these syntactic
categories \emph{modes}, and modalities will map formulas of one mode to
formulas in another. Modes can be conceived of as `possible universes of
discourse' in which we can make various logical statements. Modalities will then
allow formulas in one mode to appear in another---not directly, but as spectres
under a modality. All the modal operators in the logic will preserve
conjunction. Thus, their essence is one of a \emph{necessity} modality.
Extending the present approach to possibility-like modalities is an open
problem.

Instead of originating from a Kripke semantics of computational interest, our
logic comes from categorical logic. In fact, it is the logical isolate of a
multimodal Martin-L\"of Type Theory \parencite{nordstrom_1990} called MTT
\parencite{gratzer_2020,gratzer_2021}. Hence, it is presented as a proof system
in the style of Gentzen's \emph{natural deduction}
\cite{prawitz_1965,prawitz_2006}. Due to a lack of a double-negation elimination
rule the resultant logic is intuitionistic. The formulation of a classical
version of this logic as well as an associated Kripke semantics for it remain
an open problem.

\section{Mode theories}
  \label{section:mode-theories}

\subsection{Modes}

To begin presenting the logic we must presuppose a set \(\Mode\) of
\emph{modes}, with typical members \(m, n, \ldots \in \Mode\). Each of these
modes corresponds to a syntactic category, thus partitioning the formulas of the
logic. We will write
\[
  \varphi \At{m}
\]
to mean that \(\varphi\) is a formula at mode \(m\).

\subsection{Modalities}

Modalities are traditionally endo\"{o}perators of the logic: a modality \(\Box\)
maps a formula \(\varphi \At{m}\) to a formula \(\Box \varphi \At{m}\) at the
same mode. Our logic breaks with tradition by featuring modalities which map
formulas to different modes. Thus, a modality indexed by \(\mu\) applied to a
formula \(\varphi \At{n}\) at mode \(n\) may yield a formula \(\Box_\mu\,\varphi
\At{m}\) at some other mode \(m\). We will also break with tradition by writing
\(\Modify{\varphi}\) for the application of the modality indexed by \(\mu\) to
\(\varphi\), instead of the more common notation \(\Box_\mu\,\varphi\).

We will specify the fact that \(\varphi \At{n}\) implies \(\Modify{\varphi}
\At{m}\) by writing
\[
  \mu : n \to m
\]
This notation says that \emph{\(\mu\) is a modality from mode \(n\) to mode
\(m\)}. We are likely to call \(m\) and \(n\) the \emph{boundary} of the
modality.\footnote{This term has its origins in higher category theory.}

One may wonder how modal operators are to be be combined. Indeed, standard
treatments of modal logic define a \emph{modality} to be a composite of modal
operators, and demonstrate various `reduction laws' that simplify such
composites; see e.g. \textcite[\S 3]{hughes_1996}. In our case, if we have two
modalities \(\nu : o \to n\) and \(\mu : n \to m\), and a formula \(\varphi
\At{o}\) we see that
\[
  \Modify[\mu]{\Modify[\nu]{\varphi}} \At{m}
\]
In a more traditional system of modal logic we might have tried to prove that
such a formula is equivalent to a simpler formula \(\Modify[\xi]{\varphi}
\At{m}\) for some modality \(\xi : o \to m\). We will once more break with
tradition by presuming that such a modality always exists. In other words, we
will assume that for any two modalities \(\nu : o \to n\) and \(\mu : n \to m\)
there exists a \emph{composite modality} \(\mu \circ \nu : o \to m\). The rules
of our logic will eventually allow us to prove for any formula \(\phi \At{o}\) a
logical equivalence
\[
  \Modify[\mu]{\Modify[\nu]{\varphi}}
    \leftrightarrow
  \Modify[\mu \circ \nu]{\varphi}
    \At{m}
\]
In order to ensure that the composition of modalities behaves well we must
assume that it is governed by some algebraic laws. In particular, we will assume
that it is \emph{associative}: for any three composable modalities \(\xi : p \to
o\), \(\nu : o \to n\), \(\mu : n \to m\) we must have
\[
  (\mu \circ \nu) \circ \xi = \mu \circ (\nu \circ \xi) : p \to m
\]
Thus, a string of modalities will compose to a unique result. Moreover, we will
assume for each mode $m \in \Mode$ an \emph{identity modality}
\[
  1_m : m \to m
\]
which will be an identity element for the composition operator \(\circ\), so
that for each \(\mu : \nu \to \mu\) it is the case that \(1_m \circ \mu = \mu =
\mu \circ 1_n\). We will later be able to prove a logical equivalence
\(\Modify[1_m]{\varphi} \leftrightarrow \varphi \At{m}\) for any \(\varphi
\At{m}\).

Readers that have encountered category theory before will immediately recognise
that we have assumed that \(\Mode\) is not just a set, but a category. Between
any two modes \(m, n \in \Mode\) (the \emph{objects} of the category) we are
given a set \(\Hom[\Mode]{m}{n}\) of modalities from \(m\) to \(n\) (the
\emph{morphisms} of the category with source \(m\) and target
\(n\)). Moreover, for any three modes $m, n, o \in \Mode$ we are given an
indexed binary operation
\[
  \circ_{m, n, o} : \Hom[\Mode]{n}{m} \times \Hom[\Mode]{o}{n} \to \Hom[\Mode]{o}{m}
\]
which is associative and has `indexed' identity elements $1_m \in
\Hom[\Mode]{m}{m}$. Thus, modes and modalities form a category, i.e. a `typed'
monoid, whose elements (morphisms) have a `source' and `target' type, and where
monoid multiplication (composition) can only happen when these types align. The
structure of a category underlies a large part of modern algebra and
mathematics. For an introduction we refer the reader to books by
\textcite{awodey_2010} and \textcite{mac_lane_1978}.

It is instructive to try to encode a very simple modal syntax as a mode theory.
Recall that traditional modal logics assume a single-mode syntax. Thus, we
define the set \(\ModeL{\K{}} = \{\bullet\}\) to consist of a unique mode
\(\bullet\). Next, we can generate the morphisms by stipulating that \(\Box :
\bullet \to \bullet\) is an endomodality on that unique mode. We can then
generate the \emph{free category} based on this data. This is essentially the
free monoid on a set of generating morphisms, subject to the restriction that in
any string of morphisms the target of a morphism always matches the source of
the next. As this happens trivially in our case (we have a unique mode), the set
of morphisms is exactly the free monoid on one generator: its elements consist
of the modalities \(\Box^n : \bullet \to \bullet\) for each \(n \in
\mathbb{N}\). The composite of two morphisms is
\[
  \Box^a \circ \Box^b = \Box^{a + b}
\]
Finally, the identity morphism for this operation is \(\Box^0\).

This generates a syntax with an infinite set of modalities: if \(\varphi
\At{\bullet}\) then
\[
  \Modify[\Box^0]{\varphi},\
  \Modify[\Box]{\varphi},\
  \Modify[\Box^2]{\varphi},\ \ldots \At{\bullet}
\]
are all well-formed formulas at mode \(\bullet\). We will see later that the
logic generated here is essentially (an intuitionistic variant of) the smallest
normal modal logic \K{} \parencite[\S 1.6]{blackburn_2001}.

\subsection{Transformations between modalities}

This technology does not suffice to encode richer settings. For example, the
\axiom{4} axiom
\[
  \Box \phi \to \Box \Box \phi
\]
is one of the two characteristic axioms of the modal logic \SFour{}
\parencite[\S 3]{hughes_1996}. We would ideally like to be able to encode this
as part of the structure of the mode theory \(\Mode\). However, none of the
`moving parts' of \(\Mode\) allows the representation of such information.

Consequently, to encode implications such as the above we will need to add
another layer to the mode theory \(\Mode\). We will postulate that between any
two `parallel' modalities \(\mu, \nu : n \to m\) with the same source and target
mode there exists a set of \emph{transformations}
\[
  \alpha : \mu \To \nu
\]
These transformations---typically denoted by the letters \(\alpha\), \(\beta\),
\ldots---encode implications between modalities. 
They may be illustrated pictorially this way:
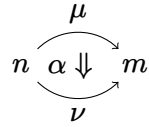
\begin{center}
  \begin{tikzpicture}
    \node (b) at (1.5,1) {$n$};
    \node (c) at (3,1) {$m$};
    \node at (2.15,1) {$\alpha \Downarrow$};
    \draw[->] (b) to [bend left=45] node[above]{$\mu$} (c);
    \draw[->] (b) to [bend right=45] node[below]{$\nu$} (c);
  \end{tikzpicture}
\end{center}
This shape is often called a \emph{globe}. We can imagine $\alpha$ as
`inhabiting' this globe. Due to this shape we are likely to collectively call
the modes \(m\), \(n\) and the modalities \(\mu\) and \(\nu\) the
\emph{boundary} of \(\alpha\).

The presence of such a transformation in \(\Mode\) will allow us to prove the
formula
\[
  \Modify[\mu]{\varphi} \to \Modify[\nu]{\varphi} \At{m}
\]
in the logic, for any formula $\varphi \At{n}$. For example, if in
$\ModeL{\K{}}$ we postulate a transformation
\[
  \axiom{4} : \Box \To \Box^2
\]
which corresponds to the \axiom{4} axiom, then in the logic we will be able to
prove the implication
\[
  \Modify[\Box]{\varphi} \to \Modify[\Box^2]{\varphi} \At{\bullet}
\]
Combined with the equivalence \(\Modify[\Box^2]{\varphi} \leftrightarrow
\Modify[\Box]{\Modify[\Box]{\varphi}} \At{\bullet}\) this implication enables a
proof of a formula that looks like axiom \axiom{4} within the logic.

The addition of \axiom{4} to a modal logic may have far-reaching implications.
For example, when combined with the \axiom{K} axiom it allows us to prove the
implication \(\Box \Box A \to \Box\Box\Box A\). Thus, there should be a minimum
amount of algebra on transformations that generates these consequences. To
start, given three parallel modalities \(\mu, \nu, \xi : n \to m\) and a formula
\(\varphi \At{n}\), the desired \emph{hypothetical syllogism}
\[
  \inferrule{
    \Modify[\mu]{\varphi} \to \Modify[\nu]{\varphi} \At{m} \\
    \Modify[\nu]{\varphi} \to \Modify[\xi]{\varphi} \At{m}
  }{
    \Modify[\mu]{\varphi} \to \Modify[\xi]{\varphi} \At{m}
  }
\]
can be indirectly encoded by the existence of a composition operation on
transformations: if \(\alpha : \mu \To \nu\) and \(\beta : \nu \To \xi\) then
there should exist a composite transformation
\[
  \beta \circ \alpha : \mu \To \xi
\]
This may be illustrated pictorially by placing two globes on top of each other:
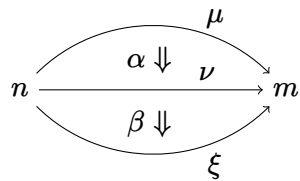
\begin{center}
  \begin{tikzpicture}
    \node (b) at (1.5,1) {$n$};
    \node (c) at (5,1) {$m$};
    \node at (3.2, 1.4) {$\alpha \Downarrow$};
    \node at (3.2, 0.55) {$\beta \Downarrow$};
    \draw[->] (b) to [bend left=45] node[above, near end]{$\mu$} (c);
    \draw[->] (b) to node[above, near end]{$\nu$} (c);
    \draw[->] (b) to [bend right=45] node[below, near end]{$\xi$} (c);
  \end{tikzpicture}
\end{center}
This composition should also be subject to associativity. Moreover, there should
be an identity transformation \(1_\mu : \mu \To \mu\) for every modality \(\mu :
n \to m\). Note that we abuse the notations for composition and identities,
using them for both modalities and their transformations.

This \emph{vertical composition} of transformations is not sufficient to
construct \(\Box\Box \varphi \to \Box\Box\Box \varphi\) from the \axiom{4} axiom
\(\Box \varphi \to \Box\Box\varphi\). What is needed instead is a form of
\emph{horizontal composition}. Suppose that we have four modalities \(\mu, \nu :
n \to m\) and \(\theta, \xi : o \to n\), and transformations \(\beta : \theta
\To \xi\) and \(\alpha : \mu \To \nu\). This may be illustrated pictorially by
placing two globes one next to the other:
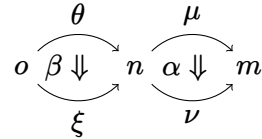
\begin{center}
  \begin{tikzpicture}
    \node (a) at (0,1) {$o$};
    \node (b) at (1.5,1) {$n$};
    \node (c) at (3,1) {$m$};
    \node at (0.6,1) {$\beta \Downarrow$};
    \node at (2.15,1) {$\alpha \Downarrow$};
    \draw[->] (a) to [bend left=45] node[above]{$\theta$} (b);
    \draw[->] (a) to [bend right=45] node[below]{$\xi$} (b);
    \draw[->] (b) to [bend left=45] node[above]{$\mu$} (c);
    \draw[->] (b) to [bend right=45] node[below]{$\nu$} (c);
  \end{tikzpicture}
\end{center}
The \emph{horizontal composition} of the transformations \(\alpha\) and
\(\beta\) is a transformation
\[
  \alpha \ast \beta : \mu \circ \theta \To \nu \circ \xi
\]
which transforms the composite modality \(\mu \circ \theta\) to the composite
modality \(\nu \circ \xi\).

If one of the two transformations is the identity then the horizontal composites
are
\begin{align*}
  1_\mu \ast \beta : \mu \circ \theta \To \mu \circ \xi
  &&
  \alpha \ast 1_\theta : \mu \circ \theta \To \nu \circ \theta
\end{align*}
This special case is sometimes called \emph{whiskering}, because its pictorial
representation resembles the adding of a cat's whisker to a transformation:
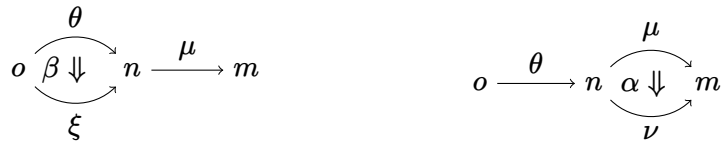
\begin{align*}
  \begin{tikzpicture}
    \node (a) at (0,1) {$o$};
    \node (b) at (1.5,1) {$n$};
    \node (c) at (3,1) {$m$};
    \node at (0.6,1) {$\beta \Downarrow$};
    \draw[->] (a) to [bend left=45] node[above]{$\theta$} (b);
    \draw[->] (a) to [bend right=45] node[below]{$\xi$} (b);
    \draw[->] (b) to node[above]{$\mu$} (c);
  \end{tikzpicture}
  &&
  \begin{tikzpicture}
    \node (a) at (0,1) {$o$};
    \node (b) at (1.5,1) {$n$};
    \node (c) at (3,1) {$m$};
    \node at (2.15,1) {$\alpha \Downarrow$};
    \draw[->] (a) to node[above]{$\theta$} (b);
    \draw[->] (b) to [bend left=45] node[above]{$\mu$} (c);
    \draw[->] (b) to [bend right=45] node[below]{$\nu$} (c);
  \end{tikzpicture}
\end{align*}
Picking \(\alpha \defeq \axiom{4} : \Box \To \Box^2\) and \(\theta \defeq \Box\)
we obtain a transformation
\[
  \axiom{4} \ast 1_\Box : \Box^2 \To \Box^3
\]
which, modulo isomorphisms, is the desired conclusion \(\Box\Box \varphi \to
\Box\Box\Box \varphi\). Thus, transformations of modalities along with their
vertical and horizontal compositions can be used to systematically encode
various interaction laws between modalities.

It may not come as a surprise that this type of structure is already well-known:
the ingredients used above are precisely the components of a (strict)
\emph{2-category}, i.e. a category which is also equipped with morphisms between
morphisms, which can be composed vertically (i.e. in the same hom-set) as well
as horizontally (between hom-sets whose source and targets match). To have the
structure of a 2-category these two compositions need to be compatible, i.e. to
obey the \emph{interchange law}: for any modalities and transformations fitting
into the diagram
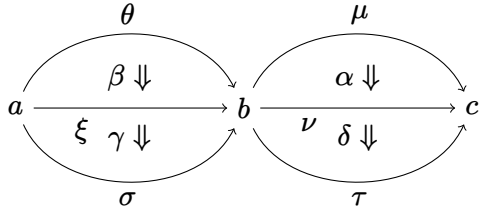
\begin{center}
  \begin{tikzpicture}[baseline=(current  bounding  box.center)]
  \node (a) at (0,0) {$a$};
  \node (b) at (3,0) {$b$};
  \node (c) at (6,0) {$c$};
  \node at (1.5,.4) {$\beta \Downarrow$};
  \node at (1.5,-.4) {$\gamma \Downarrow$};
  \node at (4.5,.4) {$\alpha \Downarrow$};
  \node at (4.5,-.4) {$\delta \Downarrow$};
  \draw[->] (a) to [bend left=65] node[above]{$\theta$} (b);
  \draw[->] (a) to node[below, near start]{$\xi$} (b);
  \draw[->] (a) to [bend right=65] node[below]{$\sigma$} (b);
  \draw[->] (b) to [bend left=65] node[above]{$\mu$} (c);
  \draw[->] (b) to node[below, near start]{$\nu$} (c);
  \draw[->] (b) to [bend right=65] node[below]{$\tau$} (c);
  \end{tikzpicture}
\end{center}
we must have that no matter which direction we compose in first, the result
should be the same:
\[
  (\delta \circ \alpha) \ast (\gamma \circ \beta)
    =
  (\alpha \ast \beta) \circ (\delta \ast \alpha)
\]

The structure of 2-categories is rich, and of foundational interest to category
theory. Of course, the terminology is different: higher category theorists do
not speak of modes, modalities, and transformations, but of morphisms and
$n$-cells. The correspondence of terms between 2-categories and our multimodal
logic can be summarised as follows:
\begin{align*}
  \text{object} &\sim \text{mode} \\
  \text{morphism (1-cell)} &\sim \text{modality} \\
  \text{2-cell} &\sim \text{transformation (natural map between modalities)}
\end{align*}
In this manner we are able to give a very precise definition of a mode theory:
\begin{definition}
  A \emph{mode theory} is a (strict) 2-category.
\end{definition}
Unfortunately, we cannot expand on the subject any further in this paper. For
introductory treatments of 2-categories we refer the reader to books by
\textcite[\S XII.3]{mac_lane_1978} and \textcite[\S 7]{borceux_1994}.

\section{Formulas and Judgements}

Having sketched how mode theories can be used to encode the modal structure of a
modal logic, we now turn to defining the formulas of our logic as well as its
proof system.

Owing to the roots of our work in Martin-L\"{o}f type theory, almost all our
definitions will be given using Martin-L\"{o}f's methodology of
\emph{judgements} \parencite{martin-lof_1996}. This amounts to a universal use
of positive statements which are inductively justified by evidence. The
canonical examples of this methodology are the proof systems of natural
deduction and sequent calculus: each sequent is a judgement, and the evidence
that a judgement holds is a proof tree with that conclusion. This methodology is
very common in the parts of Computer Science that are influenced by type theory;
see e.g. the book of Robert Harper on the foundations of programming languages
\parencite{harper_2016}. It has also been particularly influential in treatments
of the Curry-Howard correspondence for modal logic; see e.g.
\textcite{pfenning_2001}.

\subsection{Formulas}

The majority of presentations of modal logic assume a propositional syntax that
has been augmented by a set of endomodalities---usually \(\Box\) and
\(\diamond\), or an indexed version of them in the multimodal case. We will
enrich this by including a modal operator \(\Modify{-}\) for every modality
\(\mu : n \to m\) in the mode theory \(\Mode\). However, modalities transport
formulas between modes, so we have to ensure that every formula is
\emph{well-formed}. We first define a grammar of \emph{pre-formulas}. Then, we
introduce a judgement
\[
  \IsWff{\varphi}<m>
\]
which states that the pre-formula \(\varphi\) is well-formed with respect to the
mode theory \(\Mode\). Thus, the well-formed formulas of the logic are a subset
of the pre-formulas.

The \emph{pre-formulas} of are generated by the grammar
\[
  \varphi, \psi \Coloneqq
    p_i \mathrel{\Big\vert} \bot
        \mathrel{\Big\vert} \top
        \mathrel{\Big\vert} \varphi \lor \psi
        \mathrel{\Big\vert} \varphi \land \psi
        \mathrel{\Big\vert} \Impl{\varphi}{\psi}
        \mathrel{\Big\vert} \Modify{\varphi}
\]
where $\mu$ is a modality in $\Mode$. These are mostly standard. Each $p_i$ is a
propositional variable, and we have the usual propositional connectives. As is
usual in intuitionistic logic, we define \(\lnot \varphi \defeq \varphi \to
\bot\). The only deviant is the implication $\Impl{\varphi}{\psi}$, whose
antecedent carries a modality \(\mu\). Written in terms of the modal operator
and the traditional connective of implication, this is essentially
\(\Modify{\phi} \to \psi\). However, there are technical advantages in having
this compound version of implication in the logic: many proofs become
significantly shorter, and the relevant `modal modus ponens' rule is interesting
from a modal perspective. We write the usual implication \(\varphi \to \psi\)
as shorthand for \(\Impl[1]{\varphi}{\psi}\).

The \emph{well-formed formulas} (wffs) are generated by the following inductive
definition:
\begin{mathpar}
  \inferrule{
  }{
    \IsWff{p_i}
  }
  \and
  \inferrule{
  }{
    \IsWff{\True}
  }
  \and
  \inferrule{
  }{
    \IsWff{\False}
  }
  \and
  \inferrule{
    \IsWff{\varphi} \\
    \IsWff{\psi}
  }{
    \IsWff{\varphi \land \psi}
  }
  \and
  \inferrule{
    \IsWff{\varphi} \\
    \IsWff{\psi}
  }{
    \IsWff{\varphi \lor \psi}
  }
  \and
  \inferrule{
    \mu : n \to m \\
    \IsWff{\varphi}<n> \\
    \IsWff{\psi}
  }{
    \IsWff{\Impl{\varphi}{\psi}}
  }
  \and
  \inferrule{
    \IsWff{\varphi}<n> \\
    \mu : n \to m
  }{
    \IsWff{\Modify{\varphi}}<m>
  }
\end{mathpar}
With the exception of the implication and the modal operator, the rest of the
rules all refer to a single mode $m$, in which they are parametric. Thus, most
of the connectives are \emph{mode-local}: they construct propositions that
remain in a single mode. In contrast, both the rules for the modal operator and
the implication rules reach across modes. In the first case, a formula that is
well-formed at $n$ may appear in mode $m$, but only under a modality $\mu : n
\to m$. In the second case, the antecedent of an implication should be
well-formed under the appropriate modality, in a similar manner.

\subsection{Judgments and Contexts}

A \emph{judgement} of the multimodal logic has the form
\[
  \IsFm{\varphi}
\]
where $\Gamma$ is a context (at mode \(m\)), and $\varphi$ is a well-formed
formula (at mode \(m\)).

Traditionally, contexts in natural deduction consist of a list of assumptions
\(\phi_1, \dots, \phi_n\). However, in order to accommodate modal reasoning,
ours will feature two additional gadgets: \emph{tags} and \emph{locks}. Each of
these gadgets complements the other.

Each assumption in the context will be \emph{tagged} with a modality. Hence, the
assumption
\[
  \Decl[\mu]{\varphi}
\]
is meant to be read as `the formula \(\varphi\) under modality \(\mu\).' In
broad strokes this is logically equivalent to the assumption
\(\Modify{\varphi}\). When we come to define contexts we must remember to
ensure that \(\varphi\) be well-formed under \(\mu\).

The other side of the coin is the appearance of \emph{locks} in contexts. Unlike
tags, locks are operators that act on entire contexts, and are annotated by a
modality. If \(\mu : n \to m\) is a modality and \(\Gamma\) is a context at the
appropriate mode, then
\[
  \LockCx{\Gamma}
\]
will also be a context, also at an appropriate mode. We use postfix notation for
reasons that will be revealed shortly. Finally, it should be stressed that locks
are formal operations that act on the entire context; it might be perhaps more
apt to think of \(\LockCx{\Gamma}\) as \(\Lock_\mu(\Gamma)\).

As is suggested by the notation, locks restrict access to the assumptions they
enclose: whether an assumption \(\Decl[\nu]{\varphi}\) found in
\(\LockCx{\Gamma}\) shall be accessible will depend on the transformations
between modalities \(\mu\) and \(\nu\). For this reason, it is important that
contexts are understood as structures generated by a certain grammar, and not as
lists or multisets of assumptions.

In summary, the \emph{pre-contexts} are generated by the grammar
\[
  \Gamma\ \Coloneqq\
          \Emp\
    \mathrel{\Big\vert} \Gamma, \Decl[\mu]{\varphi}\
    \mathrel{\Big\vert} \LockCx{\Gamma}<\mu>
\]
where \(\Emp\) is the empty context, \(\varphi\) is a pre-formula, and \(\mu\)
is modality in \(\Mode\).

The (well-formed) \emph{contexts} are isolated by a judgement
\[
  \IsCx{\Gamma}<m>
\]
which is generated by the following rules.

\begin{mathpar}
  \inferrule{
  }{
    \IsCx{\Emp}
  }
  \and 
  \inferrule{
    \IsCx{\Gamma} \\
    \mu : n \to m \\
    \IsWff{\varphi}<n>
  }{
    \IsCx{\ECxF{\Gamma}{\varphi}}
  }
  \and 
  \inferrule{
    \IsCx{\Gamma} \\
    \mu : n \to m
  }{
    \IsCx{\LockCx{\Gamma}<\mu>}<n>
  }
\end{mathpar}

Perhaps the only unexpected detail here is that locks transport contexts
backwards along modalities: if \(\IsCx{\Gamma}\) and \(\mu : n \to m\), then
\(\IsCx{\LockCx{\Gamma}}<n>\). In categorical language we would say that the
lock operation \(\LockCx{-}\) is \emph{contravariant} in the modality \(\mu\).
The reason for this will become clear when we introduce the modal rules. The
categorical essence of it is that \(\LockCx{-}\) is in some sense a \emph{left
adjoint} to the modal operator \(\Modify{-}\), and thus must have the opposite
variance to make sense.

Finally, it is important to determine how the lock operators should interact
with the composition of modalities. Suppose that we have
\begin{align*}
  \IsCx{\Gamma}<m> &&
  \nu : o \to n &&
  \mu : n \to m
\end{align*}
The rules then allow us to construct the following context:
\begin{mathpar}
  \inferrule{
    \inferrule*{
      \inferrule*{\vdots}{\IsCx{\Gamma}<m>} \\
      \mu : n \to m
    }{
      \IsCx{\LockCx{\Gamma}}<n>
    } \\
    \nu : o \to n
  }{
    \IsCx{\LockCx{\LockCx{\Gamma}<\mu>}<\nu>}<o>
  }
\end{mathpar}
However, the mode theory also provides a composite modality \(\mu \circ \nu : o
\to m\). With respect to that modality the rules then allow us to construct the
following context:
\begin{mathpar}
  \inferrule*{
    \inferrule*{\vdots}{\IsCx{\Gamma}<m>} \\
    \mu \circ \nu : o \to m
  }{
    \IsCx{\LockCx{\Gamma}<\mu \circ \nu>}<o>
  }
\end{mathpar}
We will quotient the set of contexts, so that these two constructions will be
understood to be identical. The rationale for this choice has to do with
our earlier discussion about the equivalence between the formulas
\[
  \Modify[\mu]{\Modify[\nu]{\varphi}}
    \leftrightarrow
  \Modify[\mu \circ \nu]{\varphi}
    \At{m}
\]
for any \(\varphi \At{o}\). The proof of this equivalence will be enabled by the
fact these two contexts are syntactically interchangeable.

Hence, for any \(\IsCx{\Gamma}<m>\), \(\nu : o \to n\), \(\mu : n \to m\), and
\(\phi \At{o}\), we stipulate that
\begin{align}
  \EqCx*{\LockCx{\Gamma}<1_m>}{\Gamma}<m>
    \label{equation:ctxlockid} \\
  \EqCx*{\LockCx{\LockCx{\Gamma}}<\nu>}{\LockCx{\Gamma}<\mu \circ \nu>}<o>
    \label{equation:ctxlockcomp}
\end{align}
This last equation also reveals the reason that \(\LockCx{-}\) is best written
as a postfix operator: as it is contravariant, writing it at the end preserves
the order of symbols when composing modalities.

\subsection{Rules}

\begin{figure}
  \input{rules}
  \caption{Rules of Multimodal Logic}
  \label{figure:rules}
\end{figure}

We are now able to introduce the logical rules of the system. The complete list
is given in \cref{figure:rules}.

\paragraph{Propositional connectives}

The rules for the propositional constants and connectives \(\top\), \(\bot\),
\(\land\), and \(\lor\) are the standard rules of natural deduction. The only
difference is that they have become parametric in the mode \(\At{m}\), which
they carry from premise to conclusion. In the case of \(\lor\), the elimination
rule creates `local assumptions' as usual; but because of the structure of
contexts these need to be tagged with a modality. We pick the identity modality
\(1\), so that the rule remains completely mode-local. Therefore, the rules for
all but one of the usual propositional connectives apply in an unchanged form
within a single mode. The only exception is the compound modal implication.

\paragraph{Using assumptions}

The usual variable rule of natural deduction
\[
  \inferrule{
  }{
    \Gamma, \varphi, \Delta \vdash \varphi
  }
\]
allows us to prove a conclusion if we have already assumed it in the context.

This rule does not immediately adapt to our multimodal system. There is a sense
in which modal reasoning is largely about the \emph{control of assumptions}. The
r\^{o}le of modalities very often seems to amount to a specification of who or
which state of the world `owns' an assumption, and when we should be able to use
it. In this particular setting, the logical power of an assumption is attenuated
by the presence of a lock operator \(\LockCx{-}\). The lock stops us from using
the assumptions that it guards---unless there is a transformation that
explicitly allows it.

There are three principles that determine the behaviour of locks.

\begin{quote}
  \textbf{Principle 1}. A $\mu$-variable can escape the hold of a $\mu$-lock.
\end{quote}

In symbols, this implies that the variable rule at the very least admits the inference
\[
  \inferrule{
  }{
    \IsFm[\LockCx{\Gamma, \Decl{\varphi}}]{\varphi}<n>
  }
\]
where for $\mu : n \to m$ the formation of the context presupposes that
\[
  \IsCx{\Gamma}<m>
    \qquad
  \IsWff{\varphi}<n>
\]
If we view a lock $\Lock_\mu$ as a protector of variables, we see that it acts as a $\mu$-firewall
that only authorises $\mu$-assumptions to escape its hold. In another interpretation, the appearance
of a lock at the end of a context signifies that we are currently reasoning in a $\mu$-protected
environment, so we are entitled to access $\mu$-classified information.

As we have quotiented our contexts up to
\cref{equation:ctxlockid,equation:ctxlockcomp}, this ability of a
$\mu$-assumption to escape a $\mu$-lock should be retained even when the locks
match only up to composition. For example, given \(\nu : o \to n\) and
\(\IsWff{\varphi}<o>\) we should also be able to use the variable rule to infer
\[
  \inferrule{
  }{
    \IsFm[\LockCx{\LockCx{\ECxF{\Gamma}[\mu \circ \nu]{\varphi}}<\mu>}<\nu>]{\varphi}<o>
  }
\]
precisely because \(
  \LockCx{\LockCx{\ECxF{\Gamma}[\mu \circ \nu]{\varphi}}<\mu>}<\nu>
  =
  \LockCx{\ECxF{\Gamma}[\mu \circ \nu]{\varphi}}<\mu \circ \nu> \At{o}
\).

The second principle allows us to weaken the requirement for an exact match
between the modality and the lock:
\begin{quote}
  \textbf{Principle 2}. The transformations of $\Mode$ are `keys' for the lock.
\end{quote}
In other words, suppose that for modalities $\mu, \nu : n \to m$ we have a
transformation
\[
  \alpha : \nu \To \mu
\]
in \(\Mode\). If we interpret this to mean that the modality $\nu$ implies (or
is stronger than) the modality $\mu$, then intuition has it that $\nu$-modal
assumptions should be able to `unlock' a $\mu$-lock. In symbols:
\[
  \inferrule{
    \alpha : \nu \To \mu
  }{
    \IsFm[\LockCx{\Gamma, \Decl[\nu]{\varphi}}]{\varphi}<n>
  }
\]

The final principle is already well-known:
\begin{quote}
  \textbf{Principle 3}. The variable rule should be stable under weakening.
\end{quote}
The idea here is that weakening should be admissible independently of the
position of locks: if we have an inference in context \(\LockCx{\Gamma}\) we
should also be able to admit it in either
\(\LockCx{\ECxF{\Gamma}[\nu]{\varphi}}\) or
\(\ECxF{\LockCx{\Gamma}}[\nu']{\varphi}\) for appropriately-typed modalities
\(\nu\) and \(\nu'\). Moreover, this should only apply to tagged assumptions:
introducing a new lock should by no means be admissible! That is, if we have an
inference in context \(\Gamma\), it should not in general be possible to also
have it in \(\LockCx{\Gamma}\), as $\Lock_\mu$ might protect some of the
assumptions in $\Gamma$ by prohibiting their use.

Combining those three principles  we see that the assumption rule should
more or less function in the following manner:
\begin{enumerate}
  \item
    It should gather all the locks to the right of the relevant assumption.
  \item
    It should compose the modalities associated with each one of these locks.
  \item
    It should allow the use of an assumption whenever its tag is stronger than the
    locks that protect it, i.e. the locks to its right.
\end{enumerate}
In symbols we write
\[
  \inferrule{
    \mu : n \to m\\
    \alpha : \mu \To \Locks{\Delta}
  }{
      \IsFm[\ECxF{\Gamma}{A}, \Delta]{A}
  }
\]
where the function $\Locks{-}$ is defined by the following inductive clauses:
\begin{align*}
  \Locks{\Emp} &\defeq 1 \\
  \Locks{\ECxF{\Gamma}{A}} &\defeq \Locks{\Gamma} \\
  \Locks{\LockCx{\Gamma}} &\defeq \Locks{\Gamma} \circ \mu
\end{align*}
It is evident that this function is well-defined on contexts, for it respects
\cref{equation:ctxlockid,equation:ctxlockcomp}.

\paragraph{Locks vs. modalities}

The modal rules of the system reveal the close interaction between locks and
modal operators.

Broadly speaking, the lock operators \(\LockCx{-}\) are used to `filter' the
assumptions in the context, keeping only those that are allowed in a proof of a
formula under the modality \(\Modify{-}\). This is encoded in the introduction
rule, viz.
\[
  \inferrule{
    \mu : n \to m \\
    \IsFm[\LockCx{\Gamma}]{\varphi}<n>
  }{
    \IsFm{\Modify{\varphi}}
  }
\]
which allows us to prove the modal formula \(\Modify{\varphi}\) from the context
\(\Gamma\) exactly whenever we can prove \(\phi\) from a \(\mu\)-locked
\(\Gamma\). Thus, when trying to prove \(\Modify{\varphi}\) it suffices to prove
\(\varphi\), but with restrictions on the proof. More precisely, we are able to
use only those assumptions whose modal tag is at least as strong as \(\mu\).

The modal elimination rule
\[
  \inferrule{
    \nu : o \to n \\
    \mu : n \to m \\
    \IsFm[\LockCx{\Gamma}]{\Modify[\nu]{\varphi}}<n> \\
    \IsFm[\ECxF{\Gamma}[\mu \circ \nu]{\varphi}]{\psi}
  }{
    \IsFm{\psi}
  }
\]
is the most complicated rule of the system. Its \emph{major premise} (i.e. the
premise whose connective is being eliminated) is
\(\IsFm[\LockCx{\Gamma}]{\Modify[\nu]{\varphi}}<n>\). Notice that this judgement
could be turned into \(\IsFm{\Modify{\Modify[\nu]{\varphi}}}<m>\) by applying
the introduction rule. Putting the transformed major premise and the minor
premise side-by-side
\begin{align*}
  \IsFm{\Modify{\Modify[\nu]{\varphi}}}<m>
  &&
  \IsFm[\ECxF{\Gamma}[\mu \circ \nu]{\varphi}]{\psi}
\end{align*}
we see that this elimination rule is almost a cut rule! This is particularly
evident if we recall that \(\Modify{\Modify[\nu]{\varphi}}\) is supposed to be
logically equivalent to \(\Modify[\mu \circ \nu]{\varphi}\), which is also
supposed to be equivalent to the tagged assumption \(\Decl[\mu \circ
\nu]{\varphi}\).

Despite appearances, this elimination rule is subtle: it allows the prover to
`split' a composite modality \(\mu \circ \nu\) into its constituent parts,
keeping the second half \(\mu\) as a lock in the context of the major premise,
and eliminating only the first half \(\nu\). In fact, we will see in
\cref{section:examples} that the modal elimination rule is the central device
that allows highly non-trivial interactions between modalities to appear as
reasoning principles in the logic.

\paragraph{Implication}

As is usual in natural deduction, the implication introduction rule
\[
  \inferrule{
    \IsFm[\ECxF{\Gamma}{\varphi}]{\psi}
  }{
    \IsFm[\Gamma]{\Impl[\mu]{\varphi}{\psi}}
  }
\]
internalises the usual deduction theorem as a rule of the proof system, by
allowing the prover to discharge an assumption. This is exactly why the compound
implication \(\Impl{\varphi}{\psi}\) is a natural connective in this logic: its
antecedent mirrors the structure of assumptions in the proof system.

The elimination rule is a form of \emph{modal modus ponens}:
\[
  \inferrule{
    \mu : n \to m \\
    \IsFm{\Impl[\mu]{\varphi}{\psi}}<m> \\
    \IsFm[\LockCx{\Gamma}]{\varphi}<n>
  }{
    \IsFm{\psi}
  }
\]
If we can prove the implication \(\Impl{\varphi}{\psi}\) then proving
\(\varphi\) in a \(\mu\)-locked context suffices to obtain \(\psi\). Notice once
more that the minor premise can be transformed into
\(\IsFm{\Modify{\varphi}}<m>\) by one application of the modal introduction
rule. Thus, if we consider the assumption \(\Decl{\varphi}\) and the formula
\(\Modify{\varphi}\) to be equivalent, this rule is simply modus ponens, but a
little bit more accommodating towards the structure of locks.

\subsection{Metatheory}
  \label{section:logic-metatheory}

The system satisfies a number of the usual metatheorems. First, one is able to
show the admissibility of the usual structural rules of weakening and exchange.
Some additional care is needed in the case of weakening to ensure that the
weakened context is well-formed.

\begin{theorem}[Structural rules]
  \label{theorem:structf}
  The following rules are admissible.
  \begin{mathpar}
    \inferrule{
      \IsCx{\Gamma, \Decl[\mu]{\varphi}, \Delta}<p> \\
      \IsFm[\Gamma, \Delta]{C}<p>
    }{
      \IsFm[\Gamma, \Decl[\mu]{\varphi}, \Delta]{C}<p>
    }
    \and 
    \inferrule{
      \IsFm[\Gamma, \Decl[\mu]{\varphi}, \Decl[\nu]{\psi}, \Delta]{C}<p>
    }{
      \IsFm[\Gamma, \Decl[\nu]{\psi}, \Decl[\mu]{\varphi}, \Delta]{C}<p>
    }
  \end{mathpar}
\end{theorem}

We cannot in general weaken a context by adding a lock. In fact, locks transport
contexts between modes, so adding arbitrary locks to a context may well map a
well-formed context $\IsCx{\Gamma}$ to one that is not well-formed. However, we
can `weaken a \(\mu\)-lock' by replacing it with one corresponding to a
\(\nu\)-lock for a `weaker' \(\nu\), i.e. a modality with the same boundary
(source and target modes) for which there exists some \(\alpha : \mu \To \nu\).

\begin{theorem}[Lock Weakening]
  \label{theorem:lockwkf}
  The following rule is admissible.
  \[
    \inferrule{
      \IsFm[\LockCx{\Gamma}, \Delta]{\varphi}<p> \\
      \alpha : \mu \To \nu
    }{
      \IsFm[\LockCx{\Gamma}<\nu>, \Delta]{\varphi}<p>
    }
  \]
\end{theorem}

\noindent Finally, we can prove that a modal version of the cut rule is
admissible.

\begin{theorem}[Cut]
  \label{theorem:cutf}
  The following rule is admissible:
  \[
    \inferrule{
      \IsFm[\LockCx{\Gamma}]{\varphi}<n> \\
      \IsFm[\Gamma, \Decl{\varphi}, \Delta]{\psi}<b>
    }{
      \IsFm[\Gamma, \Delta]{\psi}<b>
    }
  \]
\end{theorem}

\noindent These metatheorems will follow as corollaries of theorems in
\cref{section:terms}.

\section{Examples}
  \label{section:examples}

In this section we demonstrate modal reasoning using our proof system.

Recall that \(\varphi \to \psi \defeq \Impl[1]{\varphi}{\psi}\). The usual modus
ponens is then a \emph{derived} rule:
\[
  \inferrule*{
    \IsFm{\varphi \to \psi} \\
    \IsFm{\varphi} \\
  }{
    \IsFm{\psi}
  }
\]
This follows from the elimination rule, as by \cref{equation:ctxlockid} we
have \(\LockCx{\Gamma}<1> = \Gamma\).

\paragraph{Some general theorems about modal formulas}

We begin by showing some theorems that hold irrespective of the choice of mode
theory. This determines the nature of our modalities---which are shown to
automatically preserve conjunctions---and showcases the various rules in action.

First, we can show that a modal antecedent \(\Decl{\varphi}\) implies its
corresponding modal formula. For any \(\mu : n \to m\) and
\(\IsWff{\varphi}<n>\) we have
\[
  \inferrule*{
    \inferrule*{
     \inferrule*{1_\mu : \mu \To \mu}{\IsFm[\LockCx{\Decl{\varphi}}]{\varphi}<n>}
    }{
     \IsFm[\Decl{\varphi}]{\Modify{\varphi}}
    }
  }{
    \IsFm[]{\Impl{\varphi}{\Modify{\varphi}}}
  }
\]
This proves one half of the claim that \(\Decl{\varphi}\) and
\(\Modify{\varphi}\) are equivalent. The other half cannot be shown as a
theorem, as an implication cannot have \(\Decl{\varphi}\) as a conclusion.
However, the following special case of the modal elimination rule for \(\mu
\defeq 1\)
\[
  \inferrule{
    \mu : n \to m \\
    \IsFm[\Gamma]{\Modify[\mu]{\varphi}}<m> \\
    \IsFm[\ECxF{\Gamma}[\mu]{\varphi}]{\psi}
  }{
    \IsFm{\psi}
  }
\]
(which follows because \(\LockCx{\Gamma}<1> = \Gamma\) by
\cref{equation:ctxlockid}) shows how we can `promote' a modal formula
\(\Modify{\varphi}\) and use it as an assumption \(\Decl{\varphi}\) in the
context of another proof. This can be thought as a converse to above proof.

One can also show a version of the \K{} axiom \(\Box(\varphi \to \psi) \to \Box
\varphi \to \Box \psi\), where the \(\Box\) in the conclusion is replaced by a
\(\Modify{-}\), and the two antecedents are tagged:
\[
  \inferrule*{
    \inferrule*{
      \inferrule*{1_\mu : \mu \To \mu}
        {\IsFm[\LockCx{\Decl{\varphi \to \psi}, \Decl{\varphi}}]{\varphi \to \psi}} \\
      \inferrule*{1_\mu : \mu \To \mu}
        {\IsFm[\LockCx{\Decl{\varphi \to \psi}, \Decl{\varphi}}]{\psi}}
    }{
      \IsFm[\LockCx{\Decl{\varphi \to \psi}, \Decl{\varphi}}]{\phi}
    }
  }{
    \IsFm[\Decl{\varphi \to \psi}, \Decl{\varphi}]{\Modify{\psi}}
  }
\]
Consequently all the modalities in our system are necessity-type modalities.

It is interesting to ask how one can handle this type of reasoning
\emph{without} using modal antecedents in implications, i.e. replacing
antecedents \(\Decl{\varphi}\) with antecedents \(\Decl[1]{\Modify{\varphi}}\)
with a trivial modal tag and a modal formula. Navigating the difference between
\(\Decl{\varphi}\) and \(\Modify{\varphi}\) is the domain of the modal
elimination rule. For example, we can prove that we can eliminate conjunctions
under modalities. Given \(\IsWff{\varphi, \psi}<n>\) and writing \(\Gamma \defeq
\Decl[1]{\Modify{\varphi \land \psi}}, \Decl{\varphi \land \psi}\) we have
\[
  \inferrule*{
    \inferrule*{
      \inferrule*{1_{1_m} : 1_m \To 1_m}{\IsFm[\Decl[1]{\Modify{\varphi \land \psi}}]{\Modify{\varphi \land \psi}}} \\
      \inferrule*{
        \inferrule*{
          \inferrule*{1_\mu : \mu \To \mu}{
            \IsFm[\LockCx{\Gamma}]{\varphi \land \psi}<n>
          }
        }{
          \IsFm[\LockCx{\Gamma}]{\varphi}<n>
        }
      }{
        \IsFm[\Decl[1]{\Modify{\varphi \land \psi}}, \Decl{\varphi \land \psi}]
          {\Modify{\varphi}}
      }
    }{
      \IsFm[\Decl[1]{\Modify{\varphi \land \psi}}]{\Modify{\varphi}}
    }
  }{
    \IsFm[]{\Modify{\varphi \land \psi} \to \Modify{\varphi}}
  }
\]
Notice that the modal elimination rule is used to turn the modal formula
\(\Modify{\varphi \land \psi}\) into an assumption \(\Decl{\varphi \land \psi}\)
which overpowers the \(\mu\)-lock. One can also prove the following theorems in
a similar manner:
\begin{equation}
  \label{equation:K}
  \begin{split}
    \IsFm*[]{\Modify{\varphi \to \psi} \to \Modify{\varphi} \to \Modify{\psi}} \\
    \IsFm*[]{\Modify{\varphi \land \psi} \leftrightarrow \Modify{\varphi} \land \Modify{\psi}}
  \end{split}
\end{equation}
Both of these are versions of the \K{} axiom.

\paragraph{Normality}

Most modal logics are single-mode, single-modal-operator logics. Following our
approach in \cref{section:mode-theories} we want to construct a mode theory
consisting of a single object \(\bullet\). The axioms of 2-categories then
dictate that we define a category \(\Hom[\Mode]{\bullet}{\bullet}\) of
modalities and their transformations. The \emph{objects} of this category are
the modalities, and the \emph{morphisms} are the transformations between them.
There also needs to be a composition functor
\[
  \circ : \Hom[\Mode]{\bullet}{\bullet} \times \Hom[\Mode]{\bullet}{\bullet}
    \to \Hom[\Mode]{\bullet}{\bullet}
\]
On objects this functor maps any two modalities to their composite; on morphisms
it maps two transformations of modalities to their \emph{horizontal composite}.

Suppose that, as in \cref{section:mode-theories}, we define \(\ModeL{\K{}}\) to
be the free category on one generator, so that \(\Hom[\Mode]{\bullet}{\bullet}\)
is the \emph{set} consisting of the modalities \(\Box^n : \bullet \to \bullet\)
for each \(n \in \mathbb{N}\). Defining \(\Box \varphi \defeq
\Modify[\Box]{\varphi}\) the proofs of \cref{equation:K} read
\begin{align*}
  \IsFm*[]{\Box{(\varphi \to \psi)} \to \Box{\varphi} \to \Box{\psi}} \\
  \IsFm*[]{\Box{(\varphi \land \psi)} \leftrightarrow \Box{\varphi} \land \Box{\psi}}
\end{align*}
Thus the `simplest' mode theory \(\ModeL{\K{}}\) generates a logic that is a lot
like \K{}.

\paragraph{Axioms as transformations}

We will now demonstrate how the transformations of the mode theory gives rise to
theorems that are usually axioms of normal modal logics.

To add axioms to the logic we can then promote the set
\(\Hom[\Mode]{\bullet}{\bullet}\) itself to be the free category on additional
transformations. If we also freely add horizontal composites we get a \emph{free
2-category}. For example, if as in \cref{section:mode-theories} we generate the
free 2-category on
\[
  \axiom{4} : \Box \To \Box^2
\]
then we get a category with an infinite number of transformations, e.g.
\[
  \begin{array}{lcl}
    \axiom{4}                          &:& \Box \To \Box^2 \\
    1_\Box \ast \axiom{4}              &:& \Box^2 \To \Box^3 \\
    1_\Box \ast 1_\Box \ast \axiom{4}  &:& \Box^4 \To \Box^5 \\
                                    &\vdots&
  \end{array}
\]
Axiom \axiom{4} then appears in the logic through the following proof: for any
\(\IsWff{\varphi}<\bullet>\),
\[
  \inferrule*{
    \inferrule*{
      \inferrule*{1_1 : 1 \To 1}
        {\IsFm[\Decl[1]{\Modify[\Box]{\varphi}}]{\Modify[\Box]{\varphi}}<\bullet>}
      \inferrule*{
        \inferrule*{
          \inferrule*{
            \axiom{4} : \Box \To \Box^2
          }{
            \IsFm[
              \LockCx{\Decl[1]{\Modify[\Box]{\varphi}}, \Decl[\Box]{\varphi}}<\Box^2>
            ]{\varphi}<\bullet>
          }
        }{
          \IsFm[\Decl[1]{\Modify[\Box]{\varphi}}, \Decl[\Box]{\varphi}]
            {\Modify[\Box^2]{\varphi}}<\bullet>
        }
      }{
        \IsFm[\Decl[1]{\Modify[\Box]{\varphi}}]{\Modify[\Box^2]{\varphi}}<\bullet>
      }
    }{
      \IsFm[\Decl[1]{\Modify[\Box]{\varphi}}]{\Modify[\Box^2]{\varphi}}<\bullet>
    }
  }{
    \IsFm[]{\Modify[\Box]{\varphi} \to \Modify[\Box^2]{\varphi}}<\bullet>
  }
\]
Similarly, we could have added an axiom
\[
  \axiom{T} : \Box^1 \To \Box^0
\]
which leads to the modal logic \T{}.

We would expect that combining axioms \axiom{4} and \axiom{T} generates the
modal logic \SFour{}. We can indeed generate a free category out of these two
generating transformations, but there is more subtlety involved. The reason is
that our mode theory reifies axioms as transformations---actual objects that can
be composed in more than one way. For example, we can immediately find three
transformations \(\alpha : \Box \To \Box\). One is simply the identity \(1_\Box
: \Box \To \Box\). But there are also two more, which combine the \axiom{T} and
\axiom{4} axioms:
\begin{align*}
  &(\axiom{T} \ast 1_\Box) \circ \axiom{4} : \Box \To \Box \\
  &(1_\Box \ast \axiom{T}) \circ \axiom{4} : \Box \To \Box
\end{align*}
Moreover, there are two ways to construct a transformation \(\Box \To \Box^3\):
\begin{align*}
  &(\axiom{4} \ast 1_\Box) \circ \axiom{4} : \Box \To \Box^3 \\
  &(1_\Box \ast \axiom{4}) \circ \axiom{4} : \Box \To \Box^3
\end{align*}
It is not unreasonable to postulate that these different ways of constructing
the same transformation are equal, i.e. that
\begin{gather}
  (\axiom{T} \ast 1_\Box) \circ \axiom{4} = 1_\Box = (1_\Box \ast \axiom{T}) \circ \axiom{4} \\
  (\axiom{4} \ast 1_\Box) \circ \axiom{4} = (1_\Box \ast \axiom{4}) \circ \axiom{4}
\end{gather}
In category theory such equations are called \emph{coherence equations}: they
state that multiple ways of performing a certain transformation are in fact
identical in their effect (coherent). The addition of coherence equations means
that a category is no longer freely generated.

A mode theory that satisfies these equations can be constructed explicitly: its
modalities are of the form \(\Box^n\) for \(n \in \mathbb{N}\); a transformation
\(\alpha : \Box^n \To \Box^m\) is just an order preserving function \(\alpha :
[m] \to [n]\) where \([m] \defeq \{ k \in \mathbb{N} \mid k < m \}\); and
composition of modalities is just their sum \parencite{schanuel_1986}.
Category theorists will recognise this as the \emph{walking comonad}, i.e. a
tiny 2-category \(\mathbf{Comnd}\) such that 2-functors \(\mathbf{Comnd}
\longrightarrow \mathbf{Cat}\) classify all categories equipped with a specific
comonad. The fact that this kind of object occurs in category theory provides
external justification for why the above list of equations might be seen as exhaustive.

Of course, this could be seen as being far more work than necessary. We could
have constructed a mode theory \(\ModeL{\KFour{}}^\text{idem}\) with one mode
\(\bullet\), and one modality \(\Box : \bullet \to \bullet\) that satisfies the
equation
\[
  \Box \circ \Box = \Box
\]
and no non-identity transformations. In this mode theory there is a unique
transformation \(\alpha : \Box \To \Box \circ \Box\): because the boundaries of
this transformation are equal, it is just the identity transformation \(1_\Box\)
on \(\Box\). With this mode theory we can prove a theorem corresponding to
\axiom{4}:
\[
  \inferrule*{
    \inferrule*{
      \inferrule*{1_1 : 1_\bullet \To 1_\bullet}{\IsFm[\Decl[1]{\Modify[\Box]{\varphi}}]
        {\Modify[\Box]{\varphi}}<\bullet>}  \\
      \inferrule*{
        \inferrule*{
          \inferrule{1_\Box : \Box \To \Box \circ \Box}{
            \IsFm[\LockCx{\LockCx{\Decl[1]{\Modify[\Box]{\varphi}}, \Decl[\Box]{\varphi}}<\Box>}<\Box>]
              {\varphi}<\bullet>
          }
        }{
          \IsFm[\LockCx{\Decl[1]{\Modify[\Box]{\varphi}}, \Decl[\Box]{\varphi}}<\Box>]
            {\Modify[\Box]{\varphi}}<\bullet>
        }
      }{
       \IsFm[\Decl[1]{\Modify[\Box]{\varphi}}, \Decl[\Box]{\varphi}]
        {\Modify[\Box]{\Modify[\Box]{\varphi}}}<\bullet>
      } \\
    }{
     \IsFm[\Decl[1]{\Modify[\Box]{\varphi}}]{\Modify[\Box]{\Modify[\Box]{\varphi}}}<\bullet>
    }
  }{
    \IsFm[]{\Modify[\Box]{\varphi} \to \Modify[\Box]{\Modify[\Box]{\varphi}}}<\bullet>
  }
\]
where the leaf on the right branch works exactly because \(\Box \circ \Box =
\Box\). This mode theory generates a version of the logic \KFour{}, which
combines the \axiom{K} and \axiom{4} axioms. We can also scale it to \SFour{} by
adding a transformation \(\varepsilon : \Box \To 1_\bullet\) from the \(\Box\)
modality to the identity modality. This leads to the mode theory
\(\ModeL{\SFour}^\text{idem}\), a more explicit description of which is the
following: there is one mode \(\bullet\), and the hom-category
\(\Hom[\Mode]{\bullet}{\bullet}\) consists of two objects \(\Box\) and
\(1_\bullet\) with a single morphism \(\epsilon : \Box \To 1_\bullet\) between
them.

At this point it still appears as if the mode theory \(\ModeL{\SFour{}}\)
generates almost exactly the same logic as the appreciably simpler mode theory
\(\ModeL{\SFour{}}^\text{idem}\). Modulo syntactic differences---e.g. that
\(\Modify[\Box^2]{\varphi}\) is the same as \(\Modify[\Box]{\varphi}\)---this is
true up to provability of formulas: the logic generated by this mode theory is
indeed equivalent to (an intuitionistic variant of) \SFour{} at the level of
provable theorems. However, at the level of \emph{proofs}, the logics generated
by \(\ModeL{\SFour}\) and \(\ModeL{\SFour}^\text{idem}\) are rather different!
The reasons for that are easily understood only when we use the
proofs-as-programs perspective of the Curry-Howard correspondence to study the
dynamic behaviour of proofs. For category theorists we will simply mention that
whereas $\ModeL{\SFour}$ generates a logic whose modality can be interpreted by
any comonad with a left adjoint, the mode theory $\ModeL{\SFour}^\text{idem}$
additionally requires that the said comonad be \emph{idempotent}.

\paragraph{Encoding multimodal logics}

The flexibility afforded by the mode theory means that we can encode multimodal
logics in our system. For example, we can encode a simple \emph{epistemic
logic}: if we start with a set of agents \(\mathbb{I}\), we can generate a mode
theory with a single mode \(\bullet\) and an epistemic modality \(K_i : \bullet
\to \bullet\) for each \(i \in \mathbb{I}\) (read as ``agent \(i\)
knows'')~\parencite[\S 12]{benthem_2010}. If we then add enough
transformations---as above---we can capture two of the most popular axioms of
epistemic logic:
\begin{align*}
  &K_i \varphi \to \varphi           &&\text{veridicality} \\
  &K_i \varphi \to K_i (K_i \varphi) &&\text{positive introspection}
\end{align*}
The axiom \(\lnot K_i \varphi \to K_i \lnot K_i \varphi\) of \emph{negative
introspection} cannot be captured as negation is not a modality in our system
(it cannot be: modalities preserve conjunctions).

To capture a basic \emph{doxastic logic}~\parencite[\S 13]{benthem_2010} we
could also add endomodalities \(B_i\) (read ``agent $i$ believes'') along with a
transformation
\[
  \textsf{Aristotle} : K_i \To B_i
\]
which states that knowledge implies belief. We could also add a \emph{strong
introspection} transformation, that states that an agent knows what they
believe:
\[
  \textsf{Introsp} : B_i \To K_i \circ B_i
\]
Whether any coherence laws naturally arise in this setting is yet to be
determined.

\paragraph{A multimode logic}

Our discussion would not be complete without including a bona fide
\emph{multimode} logic. To our knowledge no such logics have appeared before.
However, in our work on multimodal Martin-L\"{o}f type theory we have found
multimode settings extremely useful, especially when there are two distinct
`universes of discourse' that we are trying to model. The scenario usually
involves a universe of discourse in which some particular principle holds (e.g.
some axiom or induction principle), and another in which it does not. These are
related by modalities, so that the formulas in one are available in the other
under a modality, and can also be related to the formulas of another mode.

We wish illustrate that perspective in the simplest possible way. Consider the
mode theory consisting of two objects, \(\IntMode\) and \(\ClMode\), and a
single modality
\[
  \IntProv{} : \IntMode \to \ClMode
\]
The idea is that the mode \(\ClMode\) corresponds to classical logic, and the
mode \(\IntMode\) corresponds to intuitionistic logic. In this setup we are able
to add the excluded middle axiom to the rules of the classical mode:
\[
  \inferrule{
    \IsWff{\varphi}<\ClMode>
   }{
    \IsFm{\varphi \lor \lnot \varphi}<\ClMode>
  }
\]
We do \emph{not} include this rule in the logic of the intuitionistic mode
\(\IntMode\). If we can prove \(\IsFm[]{\Modify[\IntProv{}]{\varphi}}<\ClMode>\)
then we know that \(\varphi\) is a theorem of intuitionistic propositional
logic. Thus, only the theorems of intuitionistic logic are available under the
modality \(\IntProv{}\).

Notice that this modality \(\IntProv{}\) is not really an `inclusion.' For
example, we are not able to prove \(\Modify[\IntProv{}]{\varphi} \to \varphi
\At{\ClMode}\). In fact, this formula need not even be well-formed! To form
\(\IsWff{\Modify[\IntProv{}]{\varphi}}<\ClMode>\) we must have that
\(\IsWff{\varphi}<\IntMode>\), and concluding that \(\IsWff{\varphi}<\ClMode>\)
from that assumption is a non-trivial metatheorem about the logic.

In the classical mode we may infer that
\[
  \inferrule{
    \IsWff{\varphi}<\IntMode>
   }{
    \IsFm{\Modify[\IntProv]{\varphi} \lor \lnot \Modify[\IntProv]{\varphi}}<\ClMode>
  }
\]
That is: in the classical mode we can infer that it is either true or false that
\(\phi\) is intuitionistically true or false. Thus, the classical mode of this logic
can be seen as a place where one may reason about truth in intuitionistic
logic! Alternatively, the modality $\Modify[\IntProv{}]{-}$ can be seen as internalising some metatheoretic notion of provability, or even a translation from intuitionistic to classical logic.

\section{A Multimodal \texorpdfstring{$\lambda$}{lambda}-calculus}
  \label{section:terms}

In this section we establish a \emph{Curry-Howard correspondence}
\parencite{howard_1980,girard_1989,gallier_1993,sorensen_2006} for multimodal
logic. Curry-Howard correspondences are traditionally achieved as follows. Beginning with a natural
deduction system, we associate \emph{variables} with assumptions of the logic.
Then, we assign a \emph{term} to each derivation. The terms themselves are
linearly-written representations of proof trees, to which they correspond
bijectively. This process is sometimes called \emph{term assignment}.

If we annotate proof trees with terms, then we can view
\begin{itemize}
  \item terms as computer programs
  \item formulas as the types of programs
  \item proof reduction as computation
\end{itemize}
In this setting the introduction and elimination rules for implication strongly
resemble functional abstraction and function application. Thus, the system of
proof terms is often a \(\lambda\)-calculus, and proof simplification can be
seen as a \emph{dynamics} of these proofs.

First, we  describe the types of our system. These are exactly the same as the
formulas, but we consistently replace \(\varphi, \psi, \dots\) with \(A, B,
\dots\), \(\land\) with \(\times\), and \(\lor\) with \(+\). The
\emph{pre-types} of are generated by
\[
  A, B \Coloneqq
    p_i \mathrel{\Big\vert} \bot
        \mathrel{\Big\vert} \top
        \mathrel{\Big\vert} A + B
        \mathrel{\Big\vert} A \times B
        \mathrel{\Big\vert} \Impl{A}{B}
        \mathrel{\Big\vert} \Modify{A}
\]

The \emph{types} are generated by the following judgement.
\begin{mathpar}
  \inferrule{
  }{
    \IsType{p_i}
  }
  \and
  \inferrule{
  }{
    \IsType{\True}
  }
  \and
  \inferrule{
  }{
    \IsType{\False}
  }
  \and
  \inferrule{
    \IsType{A} \\
    \IsType{B}
  }{
    \IsType{A \times B}
  }
  \and
  \inferrule{
    \IsType{A} \\
    \IsType{B}
  }{
    \IsType{A + B}
  }
  \and
  \inferrule{
    \mu : n \to m \\
    \IsType{A}<n> \\
    \IsType{B}
  }{
    \IsType{\Impl{A}{B}}
  }
  \and
  \inferrule{
    \IsType{A}<n> \\
    \mu : n \to m
  }{
    \IsType{\Modify{A}}<m>
  }
\end{mathpar}
Second, we need to describe the \emph{contexts} of the type system. These are
again the same as the natural deduction system, but with the addition of a
unique variable for each assumption. Contexts are generated by the rules
\begin{mathpar}
  \inferrule{
  }{
    \IsCx{\Emp}
  }
  \and 
  \inferrule{
    \IsCx{\Gamma} \\
    \IsType{A}<n> \\
    \mu : n \to m
  }{
    \IsCx{\ECx{\Gamma}{x}{A}}
  }
  \and 
  \inferrule{
    \IsCx{\Gamma} \\
    \mu : n \to m
  }{
    \IsCx{\LockCx{\Gamma}<\mu>}<n>
  }
\end{mathpar}
considered as before subject to \cref{equation:ctxlockid,equation:ctxlockcomp}.
A point of order: when we add a new binding to a context, we assume that no
other assumption uses the same variable. This allows us to uniquely identify
which assumption is being used in a proof term without any confusion.

We extend the definition of $\Locks{-}$ to cover variables in the obvious way:
\begin{align*}
  \Locks{\Emp} &\defeq 1 \\
  \Locks{\ECx{\Gamma}{x}{A}} &\defeq \Locks{\Gamma} \\
  \Locks{\LockCx{\Gamma}} &\defeq \Locks{\Gamma} \circ \mu
\end{align*}
This operation clearly preserves \cref{equation:ctxlockid,equation:ctxlockcomp},
and is hence well-defined on contexts. One can show by induction on pre-contexts
that this operation is a homomorphism with respect to concatenation, i.e. that
\[
  \Locks{\Gamma, \Delta} = \Locks{\Gamma} \circ \Locks{\Delta}
\]
when both sides are defined.\footnote{Recall that concatenation is in general
not an admissible rule of the judgment $\IsCx{\Gamma}$, as locks may interfere
with the mode $m \in \Mode$.}

\begin{figure}
  \input{terms}
  \caption{Terms of Multimodal Logic}
  \label{figure:terms}
\end{figure}

The term assignment system for multimodal logic is given in \cref{figure:terms}.
The basic judgement is of the form \(\IsTm{M}{A}\), which means that \(M\) is a
\emph{term} of type \(A\) under the context \(\Gamma\), in mode \(m\).

The typing rules closely correspond to the rules of the logic in
\cref{figure:rules}. For example, we have replaced conjuction \(\land\) by the
Cartesian product \(\times\). We may construct a proof \(\Tm{\Pair{M}{N}}\) of
\(A \times B\) by pairing together a proof \(\Tm{M}\) of \(A\) and \(\Tm{N}\) of
\(B\). Hence, the Curry-Howard correspondence is readily apparent.

One subtle point is that the terms for the introduction of an implication, the
elimination of a disjunction, and the elimination of modal term all create
\emph{bound variables}. For example, the variable \(\Tm{x}\) is bound in the
subterm \(\Tm{P}\) within \(\Case{M}{x}{P}{y}{Q}\). Similarly, the variable
\(\Tm{x}\) is bound in \(\Tm{N}\) within \(\Let[\mu][\nu]{x}{M}{N}\). Thus, the
usual rules of capture avoidance need to be employed.

\subsection{Metatheory}

We have the following metatheoretic results on the term assignment system. The
proofs of these are ordinary inductions, but require care in propagating the
various modal contraptions within terms.

It is also worth noting that any metatheorem we establish about this system is
also a metatheorem about the logic given in \cref{figure:rules}: all we have to
do is \emph{erase} the new ingredients (terms, variables, and so on). Thus, the
theorems established in this section directly correspond to the claims in
\cref{section:logic-metatheory}.

\begin{theorem}[Structural rules]
  \label{theorem:struct}
  The following rules are admissible.
  \begin{mathpar}
    \inferH{VarWk}{
      \IsCx{\Gamma, \DeclVar{x}[\mu]{A}, \Delta}<p> \\
      \IsTm[\Gamma, \Delta]{M}{C}<p>
    }{
      \IsTm[\Gamma, \DeclVar{x}[\mu]{A}, \Delta]{M}{C}<p>
    }
    \and 
    \inferH{Exch}{
      \IsTm[\Gamma, \DeclVar{x}[\mu]{A}, \DeclVar{y}[\nu]{B}, \Delta]{M}{C}<p>
    }{
      \IsTm[\Gamma, \DeclVar{y}[\nu]{B}, \DeclVar{x}[\mu]{A}, \Delta]{M}{C}<p>
    }
  \end{mathpar}
\end{theorem}

As discussed in \cref{section:logic-metatheory}, we cannot be cavalier with
adding locks to the context.  The following rule describes how to weaken already
extant locks. Given a $2$-cell $\alpha$ and two (disjoint) pre-contexts $\Gamma$
and $\Delta$, we define the \emph{partial} metatheoretic operation
\[
  \TmKey{\Tm{M}}[\Gamma][\alpha][\Delta]
\]
by the following clauses:
\begin{align*}
  \TmKey{\Tm{\Var{x}[\alpha']}}[\ECx{\Gamma}{x}[\rho]{A}, \Gamma'][\alpha][\Delta]
    &\defeq \Tm{
               \Var{x}[
                 \DelimPrn{1_{\Locks{\Gamma'}} \ast \alpha \ast 1_{\Locks{\Delta}}}
                   \circ \alpha'
               ]
             } \\
  \TmKey{\Tm{\Var{x}[\alpha']}}[\Gamma][\alpha][\ECx{\Delta}{x}[\rho]{A}, \Delta']
    &\defeq \Tm{\Var{x}[\alpha']} \\
  \TmKey*{\Lam{x}[\xi]{M}}[\Gamma][\alpha][\Delta]
    &\defeq \Tm{\Lam{x}[\xi]{\TmKey{M}[\Gamma][\alpha][\ECx{\Delta}{x}[\xi]{A}]}} \\
  \TmKey*{\App{M}{N}[\xi]}[\Gamma][\alpha][\Delta]
    &\defeq
      \App*{
        \TmKey{M}[\Gamma][\alpha][\Delta]
      }{
        \TmKey{N}[\Gamma][\alpha][\LockCx{\Delta}<\xi>]
      }[\xi] \\
  \TmKey{\MkBox[\xi]{M}}[\Gamma][\alpha][\Delta]
    &\defeq \MkBox[\xi]{\TmKey{M}[\Gamma][\alpha][\LockCx{\Delta}<\xi>]} \\
  \TmKey{\Pair{M}{N}} &\defeq \Pair{\TmKey{M}}{\TmKey{N}} \\
  \TmKey{\Proj{M}} &\defeq \Proj{\TmKey{M}} \\
  \TmKey{\Inj{M}} &\defeq \Inj{\TmKey{M}} \\
\end{align*}
\begin{align*}
  &\TmKey*{\Let[\rho][\xi]{x}{M}{N}}[\Gamma][\alpha][\Delta] \\
    &\qquad \defeq \Let[\rho][\xi]{x}{
              \TmKey{M}[\Gamma][\alpha][\LockCx{\Delta}<\rho>]
            }{
              \TmKey{N}[\Gamma][\alpha][\ECx{\Delta}{x}[\rho \circ \xi]{A}]
            } \\
  &\TmKey{\Case{M}{x}{P}{y}{Q}}  \\
    &\qquad \defeq{} \Case{\TmKey{M}}
              {x}{\TmKey{P}[\Gamma][\alpha][\ECx{\Delta}{x}[1]{A}]}
              {y}{\TmKey{Q}[\Gamma][\alpha][\ECx{\Delta}{y}[1]{B}]}
\end{align*}



\begin{theorem}[Lock Weakening]
  \label{theorem:lockwk}
  In the following rule the term in the conclusion is well-defined when the premises hold, and the
  rule itself is admissible.
  \begin{mathpar}
    \inferH{LockWk}{
      \IsTm[\LockCx{\Gamma}, \Delta]{M}{A}<p> \\
      \alpha : \mu \To \nu
    }{
      \IsTm[\LockCx{\Gamma}<\nu>, \Delta]{\TmKey{M}[\Gamma][\alpha][\Delta]}{A}<p>
    }
  \end{mathpar}
\end{theorem}

With lock weakening at hand, we define a metatheoretic operation
\[
  \Sb{N}{M/x}
\]
which stands for the \emph{substitution} of $M$ for the variable $x$ under
context $\Gamma$. In most cases this operation simply recurses appropriately
through the structure of the term. The novel clauses are
\begin{align*}
  \Sb{\Var{x}[\alpha]}{M/x}
    &\defeq \TmKey{M}[\Gamma][\alpha][\Emp] \\
  \Sb{\MkBox[\xi]{N}}{M/x}
    &\defeq \MkBox[\xi]{\Sb{N}{M/x}} \\
  \Sb*{\Let[\rho][\xi]{y}{N_0}{N_1}}{M/x}
    &\defeq \Let[\rho][\xi]{y}{\Sb{N_0}{M/x}}{\Sb{N_1}{M/x}} \\
\end{align*}
The rest are according to custom. Notice that $\Gamma$ is a global parameter to this
definition, and is only used in the base case in order to effect lock weakening.

\begin{theorem}[Cut]
  \label{theorem:cut}
  The following rule is admissible:
  \begin{mathpar}
    \inferH{Cut}{
      \IsTm[\LockCx{\Gamma}]{M}{A}<n> \\
      \IsTm[\Gamma, \DeclVar{x}{A}, \Delta]{N}{B}<b>
    }{
      \IsTm[\Gamma, \Delta]{\Sb{N}{M/x}}{B}<b>
    }
  \end{mathpar}
\end{theorem}

\paragraph{Equational theory}

With the preceding metatheorems in hand we are now able to formulate an
\emph{equational theory of terms} for this system. The equational theory
specifies a minimal set of equations between \emph{proofs} of a certain
formula/type. In particular, the cut elimination theorem suggests the
following two \emph{$\beta$-rules}:
\begin{mathpar}
  \inferrule{
    \mu : n \to m \\
    \IsTm[\ECx{\Gamma}{x}{A}]{M}{B}<m> \\
    \IsTm[\LockCx{\Gamma}]{N}{A}<n>
  }{
    \EqTm[\Gamma]{\App*{\Lam{x}{M}}{N}}{\Sb{M}[\Gamma]{N/x}}{B}
  }
  \and
  \inferrule{
    \mu : n \to m \\
    \nu : o \to n \\
    \IsTm[\LockCx{\LockCx{\Gamma}<\mu>}<\nu>]{M}{A}<o> \\
    \IsTm[\ECx{\Gamma}{x}[\mu \circ \nu]{A}]{N}{B}<m>
  }{
    \EqTm[\Gamma]{
      \Let[\mu][\nu]{x}{\MkBox[\nu]{M}}{N}
    }{
      \Sb{N}[\Gamma]{M/x}
    }{B}
  }
\end{mathpar}
A very similar equational theory was developed by
\textcite{gratzer_2020,gratzer_2021}, but for an algebraically-specified system
of dependent types.

Finally, we could also make these equations \emph{directed}, and consider them
as \emph{reductions} from one term to another. That way we could see this system
as a programming language that is equipped with an \emph{operational semantics}.

\section{Related work}

Multimode logics were inspired by the decomposition of the \(!\) modality of
Linear Logic \parencite{girard_1987} into two adjoint functors/modalities. This
was used by \textcite{benton_1995} to present Linear Logic through the LNL
(linear-non-linear) calculus, which had two modes, linear and intuitionistic.
Many years later this pattern was used by \textcite{reed_2009} in an
unpublished manuscript which presented \emph{adjoint logic}, the first multimode
and multimodal logic. The modes and modalities of the Reed's logic were
presented through a mode theory that was a preorder; in our terminology this
means that the 2-category had no transformations, and between two modes there
was at most one modality.

The 2-categorical specification of mode theories was
introduced by \textcite{licata_2016}, who presented a single-premise,
single-conclusion, multimodal sequent calculus with adjoint modalities. This was
later refined by \textcite{licata_2017} into a multimode and multimodal
framework that also subsumes a number of substructural logics.
While much of work of Reed, Licata, and collaborators concerned sequent calculi,
they did also present a natural deduction framework for a general modal type
theory. Unfortunately, the generality of the theory meant that the rules were rather complex and
involved ubiquitous annotation by modal and substructural information. This precluded their immediate
generalization to practicable modal type theories.

A decisive step towards that direction happened with the re-introduction of Fitch-style modal
\(\lambda\)-calculi by \textcite{clouston:fitch:2018}. The Fitch style of natural deduction, which
mirrors the classic opening and closing of proof boxes at the level of proof terms, was adapted to
formulate two modal Martin-L\"of type theories, one by \textcite{birkedal_2020} and one by
\textcite{gratzer_2019}. These arise from a Fitch-style formulation of \K{} and \SFour{}
respectively.

The next step, which was that of generalising modal Martin-L\"of type theories to a multimode,
multimodal setting, proved more challenging. In particular, generalizing the elimination rule proved
to be problematic. Later work would show that these elimination rule's good behavior relied on
additional structure~\parencite{gratzer_2022}. In the case of a single modality, this additional
structure was often present on the syntax as an admissiblity, but for multiple modalities it was
necessary to manually postulate. However, adding such structure explicitly imposed further
restrictions on which modalities could be incorporated into the logic, making the elimination rule
less desirable as a basis for a general framework.

A solution was given by \textcite{gratzer_2020,gratzer_2021}, who combined Reed's mode
theories with a Fitch-style `lock' operation on contexts, and an elimination
rule the dual-context style of Davies and Pfenning
\parencite{davies_2001,pfenning_2001,kavvos_2020}. This particular combination
proved to work well in practice, leading to many examples of multimodal
type-theoretic reasoning. This type theory directly inspired the logic and modal
\(\lambda\)-calculus in this paper. Unlike \emph{op. cit.} we present the
calculus in elementary terms, i.e. without using the machinery of generalised
algebraic theories.

Before the work by \textcite{gratzer_2020,gratzer_2021} there was a limited
number of type theories with multiple modalities. These were usually ad-hoc, as
the approach was almost always guided by special properties of the modalities of
interest. With no claims to completeness we mention the work of
\textcite{pfenning_2001b}, \textcite{shulman_2018}, \textcite{nuyts_2017}, and
\textcite{nuyts_2018}.

\section*{Acknowledgements}

This work was supported in part in part by a Villum Investigator grant (no.
25804), Center for Basic Research in Program Verification (CPV), from the VILLUM
Foundation. We would like to thank Andrew Hirsch for encouraging us to write
this exposition. We are also very grateful to Celia Li and Liang-Ting Chen for
their careful reading and helpful suggestions.

\printbibliography

\end{document}

%% file: rules.tex
\begin{mathpar}
  \inferrule{
    \mu : n \to m \\
    \alpha : \mu \To \Locks{\Delta}
  }{
    \IsFm[\ECxF{\Gamma}[\mu]{\varphi}, \Delta]{\varphi}<n>
  }
  \and
  \inferrule{
  }{
    \IsFm[\Gamma]{\True}
  }
  \and 
  \inferrule{
    \IsFm{\False}
  }{
    \IsFm{\varphi}
  }
  \and 
  \inferrule{
    \IsFm{\varphi} \\
    \IsFm{\psi}
  }{
    \IsFm{\varphi \land \psi}
  }
  \and 
  \inferrule{
    \IsFm{\varphi_1 \land \varphi_2}
  }{
    \IsFm{\varphi_i}
  }
  \and 
  \inferrule{
    \IsFm{\varphi_i}
  }{
    \IsFm{\varphi_1 \lor \varphi_2}
  }
  \and 
  \inferrule{
    \IsFm{\varphi \lor \psi} \\
    \IsFm[\ECxF{\Gamma}[1]{\varphi}]{C} \\
    \IsFm[\ECxF{\Gamma}[1]{\psi}]{C}
  }{
    \IsFm{C}
  }
  \and 
  \inferrule{
    \IsFm[\ECxF{\Gamma}{\varphi}]{\psi}
  }{
    \IsFm[\Gamma]{\Impl[\mu]{\varphi}{\psi}}
  }
  \and 
  \inferrule{
    \mu : n \to m \\
    \IsFm{\Impl[\mu]{\varphi}{\psi}}<m> \\
    \IsFm[\LockCx{\Gamma}]{\varphi}<n>
  }{
    \IsFm{\psi}
  }
  \and 
  \inferrule{
    \mu : n \to m \\
    \IsFm[\LockCx{\Gamma}]{\varphi}<n>
  }{
    \IsFm{\Modify{\varphi}}
  }
  \and 
  \inferrule{
    \nu : o \to n \\
    \mu : n \to m \\
    \IsFm[\LockCx{\Gamma}]{\Modify[\nu]{\varphi}}<n> \\
    \IsFm[\ECxF{\Gamma}[\mu \circ \nu]{\varphi}]{\psi}
  }{
    \IsFm{\psi}
  }
\end{mathpar}

%% file: terms.tex
\begin{mathpar}
  \inferH{var}{
    \mu : n \to m \\
    \alpha : \mu \To \Locks{\Delta}
  }{
    \IsTm[\ECx{\Gamma}{x}[\mu]{A}, \Delta]{\Var{x}[\alpha]}{A}<n>
  }
  \and 
  \inferH{pair}{
    \IsTm{M}{A} \\
    \IsTm{N}{B}
  }{
    \IsTm{\Pair{M}{N}}{A \times B}
  }
  \and 
  \inferH{proj}{
    \IsTm{P}{A_1 \times A_2}
  }{
    \IsTm{\Proj[i]{P}}{A_i}
  }
  \and 
  \inferH{lam}{
    \IsTm[\ECx{\Gamma}{x}{A}]{M}{B}
  }{
    \IsTm[\Gamma]{\Lam{x}[\mu][A]{M}}{\Impl[\mu]{A}{B}}
  }
  \and 
  \inferH{app}{
    \mu : n \to m \\
    \IsTm{M}{\Impl[\mu]{A}{B}}<m> \\
    \IsTm[\LockCx{\Gamma}]{N}{A}<n>
  }{
    \IsTm{\App{M}{N}}{B}
  }
  \and 
  \inferH{inj}{
    \IsTm{M}{A_i}
  }{
    \IsTm{\Inj{M}}{A_1 + A_2}
  }
  \and 
  \inferH{case}{
    \IsTm{M}{A + B} \\
    \IsTm[\ECx{\Gamma}{x}[1]{A}]{P}{C} \\
    \IsTm[\ECx{\Gamma}{y}[1]{B}]{Q}{C}
  }{
    \IsTm{\Case{M}{x}{P}{y}{Q}}{C}
  }
  \and 
  \inferH{mod}{
    \mu : n \to m \\
    \IsTm[\LockCx{\Gamma}]{M}{A}<n>
  }{
    \IsTm{\MkBox{M}}{\Modify{A}}
  }
  \and 
  \inferH{let}{
    \nu : o \to n \\
    \mu : n \to m \\
    \IsTm[\LockCx{\Gamma}]{M}{\Modify[\nu]{A}}<n> \\
    \IsTm[\ECx{\Gamma}{x}[\mu \circ \nu]{A}]{N}{B}
  }{
    \IsTm{\Let[\mu][\nu]{x}{M}{N}}{B}
  }
\end{mathpar}
